\documentclass{Interspeech}

\usepackage{booktabs}
\usepackage{multirow}

\interspeechcameraready

\title{Self-supervised learning of speech representations with Dutch archival data}

\author[affiliation={1}]{Nik}{Vaessen}
\author[affiliation={2,3}]{Roeland}{Ordelman}
\author[affiliation={1}]{David A.}{van Leeuwen}

\affiliation{Institute for Computing and Information Science}{Radboud University}{The Netherlands}
\affiliation{Human Media Interaction}{University of Twente}{The Netherlands}
\affiliation{}{Netherlands Institute for Sound and Vision}{The Netherlands}
\email{nvaessen@science.ru.nl, rordelman@beeldengeluid.nl, dvanleeuwen@science.ru.nl}


\begin{document}

\maketitle

\begin{abstract}
    This paper explores the use of Dutch archival television broadcast data for self-supervised learning of speech foundation models, specifically wav2vec 2.0.
    We first study data quality assumptions for pre-training, and show how music, noise and speaker overlap affect SSL convergence and downstream fine-tuning performance.
    Secondly, we explore effectively pre-processing strategies to convert the noisy broadcast dataset into a qualitative dataset for pre-training, by using Whisper and WhisperX.
    Thirdly, we compare mono-lingual and multi-lingual pre-training with equivalent amounts of data, and show that mono-lingual pre-training is more robust to out-of-domain data.
    Lastly, we achieve a state-of-the-art \texttt{LARGE} wav2vec 2.0 model for the Dutch language, by a continuation of pre-training a wav2vec 2.0 XLS-R model checkpoint with our 55\,k hour archival dataset.
    % Code and model checkpoints will be released after anonymous review.
\end{abstract}

\section{Introduction}

Speech foundation models, such as wav2vec~2.0~\cite{baevski2020wav2vec2}, HuBERT~\cite{hsu2021hubert}, and WavLM~\cite{chen2022wavlm}, see wide-spread usage in various speech technology tasks~\cite{yang2021superb}.
% These models differ in their self-supervised learning (SSL) objective function, they have in common that they 
% are all developed and tuned with clean, pre-processed, labeled dataset(s), namely Librispeech~\cite{panayotov2015librispeech} and Libri-light~\cite{kahn2020libri} for wav2vec~2.0 and HuBERT, while WavLM also uses GigaSpeech\footnote{The authors of WavLM specifically state they use a 10\,k hour subset of the 40\,k hour GigaSpeech dataset which is validated to not have utterances containing silence or noise.}~\cite{chen2021gigaspeech} and VoxPopuli~\cite{wang2021voxpopuli}.
% The nature of these datasets is predominately prepared speech, with varying microphone quality, quiet recording conditions, known (pseudo-anonymous) speaker labels, and the knowledge that a full recording often contains a single speaker. 
% Even though the dataset labels are not used during SSL pre-training, they aid in dataset pre-processing (especially single-speaker recordings), and we argue they cause implicit assumptions on the data SSL algorithms require to work well. 
There models use self-supervised learning (SSL) to learn speech representations on unlabeled data, and evidently generalize well to downstream speech tasks. 
One particular use-case of SSL is the paradigm of multi-lingual pre-training, followed by fine-tuning for a low-resource language, as introduced by the wav2vec~2.0 XLSR model \cite{conneau2021unsupervised}, but also relevant (with caveats) to the design of Whisper \cite{radford2023whisper}.
Also for the Dutch language, this paradigm of a fine-tuning (XLSR), or directly using (Whisper), a multi-lingual model, is a popular approach \cite{wang2023dutchrepresentations,bualan2024systematic}.
This is partially due to the fact that, as to our knowledge, there is only one foundation model specifically designed for the Dutch language, released by \cite{conneau2021unsupervised}.
In this work, we want to find out whether there is an inherent benefit to multi-lingual pre-training, or whether, given the same computational budget and dataset size, a Dutch mono-lingual pre-training is more beneficial for Dutch speech recognition.

In order to answer this question, we construct a 55\,k hour Dutch audio dataset for pre-training. 
For this purpose, we made use of a collection of (80\,k hours) archival Dutch television broadcast data from the Netherlands Institute for Sound and Vision.
However, pre-training on the raw collection failed. Therefore, we first perform an analysis on data assumptions and beneficial pre-processing steps.
%Based on the datasets mentioned above, 
We speculate, based on the qualities of typical pre-training datasets~\cite{panayotov2015librispeech}, that it needs to be easy to segment the dataset into sequences of audio with speech (utterances). 
These utterances must be relative short, in the order of 1 to 30 seconds, and always contain speech for the full length of the utterance. 
Moreover, the speech in an utterance must be of a single speaker.
Finally, there is limited background noise in the dataset, and no music is present, either instrumental, vocal, or a mix of both.
Unfortunately, none of these conditions immediately apply to our collection of Dutch broadcast audio, which prompts us to perform an analysis on how to effectively clean a raw dataset for speech SSL.

Concretely, we ask the following research questions regarding self-supervised speech representation learning, and in particular, the contrastive wav2vec~2.0 algorithm, which is the only SSL algorithm we focus on due to limit computational resources:

\begin{enumerate}
\item What constitutes a high quality dataset for self-supervised speech representation learning?
\item Can we construct such a dataset from television broadcast data?
\item Is pre-training on mono-lingual audio data better than pre-training on multi-lingual audio data?
\end{enumerate}

Firstly, as speculated earlier, our hypothesis on high-quality datasets is that they need to be as similar to Librispeech (LS) as possible. 
We note that most self-supervised learning methods are solely developed on LS-like data, simply due to ease of access and availability of read speech. 
However, this implies that SSL methods are overfit on a meta level, designed to work well with the particularities of LS, but not on other kinds of datasets.
Regarding the second research question, we assume that, by ensuring our broadcast data is pre-processed correctly, it will have have similar properties to LS, and therefore it can be used as a high-quality pre-training dataset.
For the last research question, we hypothesize that, when equally sized, a mono-lingual dataset leads to higher-quality speech representations in that language, compared to a multi-lingual dataset, as more variation of that specific language can be observed, such as local dialects.  

\section{Related work}

The authors of \cite{ashihara2023japanese-ssl} compare mono-lingual pre-training with Japanese data to fine-tuning multi-lingual pre-trained models, specifically, XLSR from \cite{conneau2021unsupervised}.
They show that a \texttt{BASE} wav2vec~2.0 model, pre-trained on a dataset of 500 hours of spontaneous Japanese speech, outperforms the \texttt{LARGE} XLSR model, when fine-tuned (50 hours) and evaluated (3.5 hours) on the Japanese newspaper reading dataset from \cite{itou1999jnas}.
Moreover, they show that pre-training on 200 hours of Japanese data is enough to achieve equal or better performance to XLSR on this dataset.
In \cite{meng2022sslfastspeech} aspects of data quality when applying speech representation learning is studied. 
They vary the \textit{gender}, \textit{content} and \textit{prosody} of LS for pre-training, and evaluate these models on the SUPERB benchmark~\cite{yang2021superb}.
They find that pre-training performance does \textit{not} decrease when the pre-training has a (significant) gender imbalance.
Moreover, the complexity of the audio, i.e., whether or not the dataset contains audio with a large vocabulary of words, did not affect downstream performance either.
However, they observe a significant decrease in downstream task performance when the pre-training audio is sped up, and a significant increase in performance when the pre-training audio is slowed down.  
The authors of \cite{lamyeemui2023weaklang} continue pre-training XLSR with South-African soap opera broadcast data, which is shown to be an effective method to improve ASR performance for these low-resource languages.
% , which are often used in a code-switching setting.
% The authors of \cite{jacobs2023hatespeech} build a dataset from radio broadcasts to evaluate hate speech detection in the low-resource Swahili and Wolof language. 
The authors of \cite{mateju2023swedish} train a Swedish ASR system from scratch, using multiple labeled data sources, including television news from SVT, the Swedish public broadcasting organization.
The authors of \cite{lehecka2024oralarchive} explore bi-lingual and tri-lingual pre-training on a multi-lingual oral archive dataset. 
% Their pre-trained mono-lingual models performed well, but they did not have the computational resources to scale; 
% they observed best performance with the \texttt{large-v2} Whisper model from \cite{radford2023whisper}. 

\section{Methodology}

\subsection{Pre-training and fine-tuning}

In this paper we use the contrastive learning approach of wav2vec~2.0 for pre-training speech representations.
We also fine-tune the network for automatic speech recognition with labeled data using CTC loss~\cite{graves2006ctc}. 
We do not make any modification to the network architecture, 
making use of the \texttt{BASE} (12 transformer layers) and \texttt{LARGE} (24 transformer layers) variant. 
% We do not apply LayerDrop~\cite{fan2020layerdrop} to simplify the distributed data-parallel training.
We do not change any hyperparameters for pre-training or fine-tuning, expect for the learning rate, batch size, and amount of training steps.
Pre-training hyperparameters are detailed in the experiment section.
For fine-tuning, for all conditions we use a learning rate of \num{5e-5}, and 80\,k steps, with the transformer network frozen for the first 5\,k steps, and the CNN frozen for all steps.
For fine-tuning and evaluating for English, we train on 10 minutes or 100 hours of LS \cite{panayotov2015librispeech}, and evaluate on test-clean and test-other.
For fine-tuning and evaluating for Dutch, we use the corresponding train and test splits of the Dutch subset of multi-lingual Librispeech (MLS)~\cite{pratap2020mls} and the Dutch subset of CommonVoice (CV), version~18~\cite{ardila2020commonvoice}. 
We also finetune models on CGN~\cite{Oostdijk:2003}, using the official acoustic training components defined for the ``N-Best''~\cite{N-best:2009} evaluation, a benchmark for Dutch ASR, but leaving out Conversational Telephone Speech for Southern Dutch.  As corresponding evaluation material we use part of N-Best~\cite{N-best:2009} itself, namely the Northern Dutch Broadcast News section of the evaluation, with caseless word comparison.
We normalize Dutch text such that the train and test data matches the letter vocabulary used in \cite{baevski2020wav2vec2} for English, i.e., 26 letters, apostrophe, and a space.
All other punctuation, including hyphens, is substituted with spaces.
Accented letters are converted to their form in the English alphabet.
We use letter-decoding in a greedy manner.
% , we do not use any language models.
% For Dutch,   

\subsection{Data quality simulation} \label{meth:data-quality}

We first study the effect of data quality on English self-supervised speech representation learning, by using LS~\cite{panayotov2015librispeech}. 
% This train set consists of 281\,k English utterances, with a minimum duration of 0.8 seconds, a maximum duration of 29.7 seconds, and an average duration of 12.3 seconds. 
% In total, there are 2338 unique speakers (1210 male, 1128 female), with 51.7\%/48.3\% of the utterances being spoken by a male/female speaker.
% Each utterance is read speech, sourced from volunteers recording public domain audiobooks. 
% This implies that there is little session variability for speakers, as it is unlikely volunteers would change recording conditions (microphone, room conditions, time factors such as sickness, or aging). 
We use the median length utterance in all chapters (``sessions'') of each speaker for a validation set consisting of 2\,\% (20 hours) of the train set. 
The \texttt{dev-clean} and \texttt{dev-other} set are used for choosing hyperparameters.
% and final (fine-tuning) results are reported on the \texttt{test-clean} and \texttt{test-other} set.  

We augment the 960 hour train set of LS in various ways to simulate different data qualities. 
For these augmentations, we use MUSAN dataset~\cite{snyder2015musan} as a source of background noise samples,
% MUSAN consists of three subsets (speech, music, noise). 
% We only make use of the noise subset, which consists of 929 files with an average duration of 25 seconds, for a total of \~6 hours of data.
% There are 3 categories of noise, namely 
% 1) technical sounds, e.g., dial tones
% 2) ambient sounds, such as creaking doors and police sirens, and 3) intelligible speech, i.e., recordings of crowd noises.
and the FMA (large subset) dataset~\cite{defferrard2017fma} is used to provide music audio.
% We use the large FMA subset, which consists of 104\,k music segments which are all 30 second random chunks taken from the music recordings in the full database.
% The music has 16 top-level genres, including, e.g., classical, rock, and jazz.
Using these datasets, we simulate various forms of unclean data, which can arise when trying to pre-process a raw dataset into a pre-training dataset with the use of speech activity detection (SAD) tools. 

\textbf{Speaker overlap}
One common scenario is a dialogue between two or more individuals, where the speaker turn is nearly instantaneous, making segmentation by SAD infeasible. 
% In this case, one would use a high-quality diarization model to distinguish between speakers.
% Another scenario is that, e.g., two speakers are speaking into the same channel at the same time.
% To disentangle the audio, one would require a speech separation model. 
% However, for these simulations, we are simply interested in the effect on pre-training when these scenarios are not accounted for.
To that end, we propose three augmentations.
\texttt{1spk-cat} concatenates two random utterances from the \textit{same session} of LS, thus, all utterances contain only a single speaker. This acts as a baseline.
% where the audio is simply twice as long on average, with a minimal, but detectable, concatenation artifact, as the utterances are not in order.    
\texttt{2spk-cat} concatenates two random utterances from \textit{different speakers} of LS. This simulates a speaker turn which was not detected by diarization tools.
% We do not control for the sex of the speakers, thus, there are same-sex and different-sex speaker turns. Compared to the baseline, the utterances, similarly, includes the concatenation artifact, but also adds the necessity for the SSL method to model two speakers. 
\texttt{2spk-mix} mixes two utterances from two different speaker of LS.
This simulates the necessity of a speech separation model.
At batch creation time, we mix 10\,\% or 33\,\% of utterances in the batch.
Mixing is done with a uniformly sampled signal-to-noise ratio between 0 and 15 dB. 
% We vary the fraction to 10\,\%, 33\,\%, or 50\,\%. Note that due variable lengths of utterances, these fractions are cannot be obtained precisely, but we aim to be as close as possible, always ensuring at least one utterance is selected.  

\textbf{Music and noise}
The other scenarios we want to study is the inclusion of noise or music in utterances.
In our archival dataset the presence of these are quite common.
% For example, a broadcast could simply be a recording of a concert, documentaries usually have background music, or interviews could be conducted in a busy street. 
% To quantify the effect of these non-speech audio fragments on speech representation learning, we set-up the following six augmentations:
We propose the follow augmentations.
% \texttt{sub-noise} replaces one or more utterances in a batch with noise from MUSAN. Utterances are randomly cropped, or repeated, to match the exact length of the replaced utterance. 
\texttt{sub-noise} and \texttt{sub-music} replace one or more utterances in a batch, respectively with noise from MUSAN, or music from FMA. 
% As all segments from `FMA` are 30 seconds long, we can, in each case, randomly crop to equal length of the Librispeech utterance. 
Instead of substituting, we also mix, referred to as \texttt{mix-noise} and \texttt{mix-music}. 
% We use all three categories of noise, including crowd noises. 
% \texttt{mix-music} mixes one or more utterances in a batch with music segments from FMA. 
For music, we also explored stemming FMA, thereby mixing vocal (\texttt{mix-vocal}) and instrumental tracks (\texttt{mix-instr}). 
For stemming we used Demucs~\cite{rouard2023demucs}.
% \texttt{mix-instr} mixes one or more utterances in a batch with music segments from FMA, which were stemmed to only contain the \textit{instrumental} audio of the song track. 
% \texttt{mix-vocal} mixes one or more utterances in a batch with the music segments from FMA, which were stemmed to only contain the \textit{vocal} audio of the song track. 
% We applied a basic SAD on the vocal stem to filter out songs where were already instrumental before stemming.
For all settings, we select utterances to be mixed or replaced as done in \texttt{2spk-mix}.
% with a fraction varied between 10\,\%, 33\,\%, and 50\,\%, ensuring at least one utterance is mixed or replaced.
% For mixing, we use a signal-to-noise ratio uniformly sampled between 0 and 15. 
% Stemming of FMA into vocal and instrumental tracks was done with Demucs~\cite{rouard2023demucs}.
% and SAD on potentially silent vocal tracks was done with Pyannote~\cite{bredin2020pyannote}.

\subsection{Archival data collection}

To obtain a large quantity of Dutch audio data for self-supervised learning, we created a dataset of Dutch
% radio and
television broadcasts within a 50-year period, from 1972 to 2022, at the Netherlands Institute for Sound \& Vision (NIBG in Dutch). 
This initial dataset contained 182\,k hours of audio.
% To obtain a large quantity of Dutch audio data for self-supervised learning, we obtained television broadcasts from the Dutch public broadcasting organization (NPO) within a 50-year period, from 1972 to 2022, collected by the Netherlands Institute for Sound and Vision (NIBG).
% This initial dataset contained \~182\,k hours of audio.
First, we used meta-data to filter out genres which were unlikely to contain a good variety of speech, such as nightly news and sports broadcasting, and kept genres such as in-depth news analysis, quizzes, and documentaries.
We also removed any broadcast which lasted longer than three hours, and de-duplicated some data by heuristically removing consecutive broadcasts with the same title, summary, and publication date. 
The remaining broadcasts contained 81\,k hours of audio, with most audio files having a length of 30 minutes to 1 hour.
We refer to this 81\,k hour dataset as \texttt{nibg}
% \footnote{An identifier with meta-data will be given after peer review.}
throughout this paper.
For pre-training, short segments are required, thus, we need the data to be segmented into short utterances. 

\subsection{Segmenting the broadcast data} \label{meth:nibg-preprocess}

To perform the segmentation, we initially attempted to apply SAD with Pyannote. However, we found the resulting segmentations to contain a lot of speaker overlap, music and noise. 
Moreover, initial pre-training experiments were unsuccessful, prompting experiments with simulating data quality, as detailed above.
The second approach to segment the data involved Whisper~\cite{radford2023whisper} and WhisperX~\cite{bain2023whisperx}, with the idea that audio which can successfully transcribed by Whisper contains speech with high-enough quality to be useful for self-supervised pre-training. 
Furthermore, WhisperX combines Whisper, Wav2vec 2.0 and Pyannote to perform speaker diarization on top of the transcription, potentially leading to utterances with a single speaker, like LS. 
We experiment with 3 variations of segmenting with Whisper, and 5 variations segmenting with WhisperX.

First, we will detail our methods using the \texttt{large-v2} version of Whisper~\cite{radford2023whisper}.
The Whisper model operates on 30 seconds of audio.
Therefore, the ASR output intrinsically has a maximum length of 30 seconds.
Whisper outputs a list of segments, with a \textit{predicted} start and end-time for the spoken sentence.
The first variation, \texttt{w-raw}, simply uses these output sequences, where utterances are cut by using the indicated start and end-time in each output segment.
% However, due to the nature of transcribing consecutive chunks of 30 seconds, many utterances of a single speaker are cut between multiple segments. 
% These cuts are often indicated by \texttt{...} in the transcription.  
% Moreover, the model often transcribes non-speech fragments by describing it, with phrases like \texttt{music} or \texttt{laughing}. 
% Furthermore, the model tends to "hallucinate" on silent audio, an artifact from training on scraped, subtitled audio, which frequently has copyright information as a subtitle at the end, without paired speech.
As a second and third variation, we apply post-processing to the Whisper output in an attempt to resolve issues with mid-speech cuts, non-speech fragments, and hallucinations. 
The post-processing steps are 1) filtering out segments which exactly match certain keywords, e.g., \textit{muziek}, 2) recursively merge segments until they end with a punctuation, and 3) filter out segments shorter than 1\,s, denoted as \texttt{whisper-pp-1s}, or 3\,s, denoted as \texttt{whisper-pp-3s}.

% \begin{enumerate}

% \item Filter out any segments with transcriptions exactly matching (case-insensitive) these text phrases: \texttt{muziek}, \texttt{gelach}, \texttt{tv gelderland}, \texttt{applaus}, \texttt{gezang}, \texttt{***}, and \texttt{.} (a single dot).

% \item If the transcription of a segment does \textit{not} end with sentence-ending punctuation (a dot, question or exclamation mark), we recursively merge it with the next segment until a (merged) segment does end with such punctuation. We do not merge segments if the gap between the current and next segment is longer than 3 seconds.

% \item We remove segments which are shorter than a threshold, with the observations that many short segments are simply phrases like "yes" or "that is right"; long segments are more likely to be clean, varied speech.  
% If, due to the  merging, segments end up longer than 30 seconds, we simply cut the segment in half recursively until all subsegments are shorter than 30 seconds. 
% The two variations we use are \texttt{w-pp-1s} and \texttt{w-pp-3s}, which apply these post-processing steps on the segments from \texttt{w-raw}, with the threshold in the third step being respectively 1 and 3 seconds.

% \end{enumerate}

Secondly, we use WhisperX~\cite{bain2023whisperx}, which uses Whisper \texttt{large-v2},  Pyannote~\cite{bredin2020pyannote}, and a Dutch wav2vec 2.0 model\footnote{https://huggingface.co/jonatasgrosman/wav2vec2-large-xlsr-53-dutch}.
We use the output of each of the three stages in WhisperX inference as segmentations for pre-training.
The output of the first cut-and-merge stage is denoted as \texttt{wx-asr}.
% In the first stage, Pyannote SAD is applied to chunk the audio, with a cut and merge operation which aims to create chunks close to 30 seconds. 
% These chunk can then be transcribed by Whisper (in our case, \texttt{large-v2`} in batches, allowing for an inference speed-up. 
% Our first variation, \texttt{wx-asr}, uses the resulting output for segmenting the broadcast data.
% The second stage applies a forced alignment between the chunks and their respective transcriptions, using a wav2vec 2.0 model fine-tuned for ASR\footnote{}, after which sentences are broken up in separate segments.
The output of the second stage, which applies a forced alignment, is denoted as \texttt{wx-align}. 
The third stage applies speaker diarization, and the output is denoted as \texttt{wx-diar}.
Finally, on \texttt{wx-diar}, we also apply the post-processing techniques above. 
However, instead of merging on punctuation in the second step, we merge on speaker labels instead. We denote these \texttt{wx-diar-1s} and \texttt{wx-diar-3s}, with a threshold for 1 and 3 seconds in the third post-processing step, respectively.

\section{Experiments}

\subsection{Data quality} \label{exp:data-quality}

We pre-train the \texttt{BASE} wav2vec 2.0 variant on the 960 hours of training data from LS, but apply various augmentations as described in Section \ref{meth:data-quality}.
If not otherwise mentioned, we use the default configuration for pre-training and fine-tuning as described in \cite{baevski2020wav2vec2}.
% i.e., diversity loss scaling of $0.1$, $L^2$-penalty loss scaling of 10, gumbel-softmax temperature $tau=2$ to $tau=0.5$ with factor $0.999995$, contrastive loss temperature $tau_c=0.1$, dropout of 10\,\% throughout the network, and the feature extractor CNN gradients are scaled by a factor $0.1$.
% We also use 2 codebooks for quantization, with each codebook containing 320 entries of 128-dimensional codewords. 
% However, we do not apply LayerDrop \cite{fan2020layerdrop,huang2016layerdrop}, as this complicates distributed data-parallel training (DDP).

For pre-training, we use AdamW \cite{loshchilov2018adamw} with a weight decay of $0.1$, $\beta_1=0.9$ and $\beta_2=0.98$.
We use a batch size of 5 minutes (4.8\,M tokens) for all settings on a single NVIDIA A100 GPU. 
We train with a triangular learning rate schedule, with 25\,k linear steps up and down, and a minimum LR which is 100 times smaller than the maximum LR.
We scan for the maximum LR with a grid search between \num{e-5} and \num{e-3}.

% In the first phase we attempt LR \num{3.2e-5}, \num{1e-4} and \num{3.2e-4} with 50\,k training steps, i.e., a single cycle. 
% In the second phase we train for 400\,k steps, i.e., 8 cycles, with the best learning rate in phase 1, and 2 additional LRs, based a reasonable step size upwards or downwards, respectively:

% + If \num{3.2e-5} was best in phase 1, we also attempt LR \num{1.0e-5} and \num{1.8e-5}.
% + If \num{1e-4} was best in phase 1, we also attempt LR \num{5.6e-5} and \num{1.8e-4}
% + If \num{3.2e-4} was best in phase 1, we also attempt LR \num{5.6e-4} and \num{1.0e-3}

% The best LR is chosen according to the run with the lowest validation loss.
% The validation split, made beforehand, holds out a single utterance from each session in the training data of Librispeech. 
% We do \textit{not} apply the data quality simulation to the validation data, as we want to evaluate the quality of the speech representations on data which is similar to the fine-tuning dataset.

Fine-tuning is done with the 10 min (12\,k steps) and 100 hour (80\,k steps) labeled data conditions, using batches of 3.2\,M tokens from LS training data.
% We use the default tri-stage learning rate with a 10\,\% warm up, 40\,\% constant, and 50\,\% exponential decay phase, with the Adam optimizer \cite{kingma2015adam}, with $beta_1=0.9$, $beta_2=0.98$, and no weight decay.
We use the tri-stage LR schedule with initial LR of \num{5e-7}, a constant LR of \num{5e-5}, and a final LR of \num{2.5e-6}.
% We mask 5\,\% of the latent speech feature sequence, and use a dropout of 10\,\%.
% Inference is done greedily with letter-decoding, we do not use a language model.
% For 10 minutes of labels we train for 12\,k steps, for 100 hours of labels we fine-tune for 80\,k steps.  
% The feature CNN is frozen for the first 5\,k steps.

\begin{table}[h]
\centering
\caption{WER in \% on LS test-clean and test-other after fine-tuning various SSL data conditions.
For some conditions, x/y indicates 10/33\,\% of mix or substitute in the batch. }
\label{tab:data-quality}
\vspace{-0.9em}
\begin{tabular}{lccc}
\toprule
\multirow{2}[2]{*} & \multicolumn{1}{c}{10\,m labels} & \multicolumn{2}{c}{100\,h labels} \\
\cmidrule(lr){2-2} \cmidrule(lr){3-4}
SSL data & clean & clean & other \\
\midrule
LS baseline       & 59.6 & 12.4 & 31.5 \\
\midrule
\texttt{1spk-cat} & 62.0 & 12.8 & 32.9 \\
\texttt{2spk-cat} & 75.6 & 16.8 & 39.2 \\
\texttt{2spk-mix} & 62.5/63.9 & 12.2/13.4 & 31.9/33.4 \\
\midrule
\texttt{sub-noise} & 63.9/77.7 & 13.3/18.5 & 33.9/42.0 \\
\texttt{sub-music} & 66.5/69.9 & 14.4/15.4 & 35.3/37.5 \\
\midrule
\texttt{mix-noise} & 61.6/65.7 & 12.7/13.3 & 32.3/33.2 \\
\texttt{mix-music} & DIV       & DIV       & DIV       \\
\texttt{mix-vocal} & DIV       & DIV       & DIV       \\
\texttt{mix-instr} & 64.0/69.6 & 12.5/14.6 & 31.8/35.6 \\
\bottomrule
\end{tabular}
\end{table}

We show fine-tuning performance of all data quality simulation in Table \ref{tab:data-quality}. 
First, we see that the baseline, where we trained on the vanilla LS training dataset without any data manipulations, mostly has the best performanc.
% There is one exception, namely for \texttt{2spk-mix} the 12.2\,\% on test-clean with 100\,h labeled data is slightly better than the baseline of 12.40\,\%. 
Secondly, we see that there is a minor performance drop when concatenating utterances from the same speaker (\texttt{1spk-cat}), and a more significant performance drop for \texttt{2spk-cat}. 
Thirdly, we see that substituting 10\,\% of the data in a batch with pure noise (\texttt{sub-noise-10}) has a minor effect on performance, comparatively with concatenating the same speaker to all utterances (\texttt{1spk-cat}). 
A much larger effect is visible when the fraction of substitutions is increased to 33\,\%.
% We also observe that noise substitutions are less harmful than music substitutions in minor fractions (\texttt{sub-music-10}), but for the larger fractions, music substitutions lead to less degradation, although the performance is still significantly worse compared to `baseline`.
Fourthly, we observe that for both conditions of \texttt{mix-music} and \texttt{mix-vocal}, pre-training is not stable.
We only managed to find a converging LR when mixing instrumental music.
% For \texttt{mix-music} (which can be any combination of vocal and instrumental music), or \texttt{mix-vocal}, we see that even when mixing a small amount of the data in a batch (10\,\%) the contrastive loss diverges.
Fifthly, we see mixing in a second speaker (\texttt{2spk-mix}) only leads to relatively small performance drops. 
% The performance of \texttt{2pk-mix} sits between the performance of the baseline and \texttt{2spk-concat}. 
% Note that the impact of mixing a speaker in 50\,\% of the data is on par with substituting 10\,\% of the data with music.  

\subsection{Effective pre-processing methods} \label{exp:pre-process}

From the analysis of data quality on LS above, it seems paramount that pre-training data contains as little music as possible. To this end, we proposed the pre-processing methods for the NIBG data described in Section \ref{meth:nibg-preprocess}, with the hope that music sections will either be ignored, or transcribed as such, by the Whisper model.
As pre-processing the whole \texttt{nibg} dataset with Whisper is computationally intensive, we decided to first analyze the performance of different methods on a subset of \texttt{nibg}. 
At random, we selected 4000 broadcasts, with a total duration of 2500 hours of data. 
We applied the various pre-processing techniques to this subset.
% The resulting subsets and their statistics are displayed in Table \ref{tab:segment-statistics}.
% The \texttt{sequential} subset can be seen as a baseline with minimal pre-processing; the discard of 0.65\,\% of data is simply the truncation at the end of each file, as we discard the last segment of each file if it is not exactly 30 seconds.  
% Furthermore, we note that \texttt{wx-diar-3s} discards a lot of data, almost a third, but has the closest characteristics to Librispeech, with a minimum duration of 3 seconds for each utterance, and an average duration of 12.63 seconds.  
For each subset we perform self-supervised pre-training with hyperparameters equivalent to Section \ref{exp:data-quality}.  
First, we pre-train with all subsets, using a batch size of 5 minutes.
On the most promising subsets we also pre-train with a batch size of 40 minutes. 
Instead of a learning rate scan, we simply use the triangular schedule, with a minimum LR of \num{1e-6} and a maximum LR of \num{1e-4}.
% We selected the checkpoint with the lowest validation loss for fine-tuning. 
% As validation data we combine the validation split of Dutch MLS and Dutch CV.
% The checkpoints are fine-tuned and evaluated with the respective splits of the Dutch CV and MLS datasets.
% We describe the training splits as \texttt{MLS-nl} and \texttt{CV-nl}, and the test sets as \texttt{MLS-test} and \texttt{CV-test}.
% Fine-tuning is done with the exact same configuration as the 100\,h fine-tuning condition in Section \ref{exp:data-quality}, including the maximum learning rate of \num{5e-5}.

\begin{table}[t]
    \centering
    \caption{WER in \% when applying various pre-processing techniques on \texttt{nibg} data. We show also show average \textbf{utt}erance length of the subsets, and the batch size during pre-training.}
    \label{tab:segment-ft}
    \vspace{-0.9em}
    \begin{tabular}{lllll}
    \toprule
     subset & utt & SSL bs & MLS-test &  CV-test \\
     \midrule
     \texttt{sequential} & 30 & 5/40 & 31.0/19.7 & 39.7/17.7 \\
     \midrule
     % \texttt{w-raw} \& \texttt{w-pp-1s} & 5 & DIV & DIV \\
     \texttt{w-pp-3s}    & 4.3 & 5/40 & 25.3/16.5 & 31.3/12.8 \\
     \midrule
     \texttt{wx-asr}     & 22 & 5 & 31.3 & 39.9 \\
     \texttt{wx-align}   & 3.8 & 5 & 27.6 & 33.9 \\
     \texttt{wx-diar}    & 3.8 & 5/40 & 27.6/17.5 & 33.8/14.6 \\
     \texttt{wx-diar-1s} & 11 & 5 & 27.2 & 34.6 \\
     \texttt{wx-diar-3s} & 13 & 5/40 & 27.9/17.1 & 35.4/14.0 \\
     \bottomrule
    \end{tabular}
\end{table}

\begin{table*}[h]
\centering
\caption{WER for Dutch ASR afer fine-tuning small and large scale pre-training in mono-lingual and multi-lingual settings.
The first section only fine-tunes published pre-trained models. 
The second section sets a (small batch) baseline for mono-lingual (English or Dutch) pre-training.
The third section explores pre-training with our large mono-lingual archival dataset.}
\label{tab:large-scale-ssl}
% 14 columns
\begin{tabular}{@{}lrrllcccccccc@{}}
\toprule
& \multicolumn{4}{c}{SSL configuration} & \multicolumn{2}{c}{ft. /w MLS-nl} & \multicolumn{2}{c}{ft. /w CV-nl} & \multicolumn{1}{c}{ft. /w CGN} &  \multicolumn{3}{c}{ft. /w all three}  \\ 
\cmidrule(lr){2-5} \cmidrule(lr){6-7} \cmidrule(lr){8-9}  \cmidrule(lr){10-10} \cmidrule(lr){11-13}
network & lang & data (h) & batch & steps & MLS-t & NBest & CV-t & NBest & NBest & MLS-t & CV-t &  NBest \\
\midrule
\multicolumn{12}{l}{\textbf{Pre-trained models by Fairseq}} \\
\texttt{LARGE} \cite{baevski2020wav2vec2} & en & 1\,k & 80\,m & 400\,k & 15.9 & 45.0 & 18.5 & 38.4 & 16.5 & 17.1 &	12.1 & 18.0 \\
\texttt{LARGE} \cite{conneau2021unsupervised} & nl & 20\,k & 80\,m & 250\,k & 16.1 & 32.8 & 12.8 & 25.6 & 14.7 & 16.4 & 9.8 & 13.8 \\
\texttt{LARGE} \cite{conneau2021unsupervised} & 53 & 56\,k & 80\,m & 250\,k & 13.0 & 38.1 & 12.0 & 28.1 & 14.3 & 12.7 & 8.6 & 14.4\\
\texttt{LARGE} \cite{babu2021xlsr-large-scale} & 128 & 436\,k & 4.3\,h & 1\,M & 13.4 & 33.6 & 12.2 & 25.9 & 13.2 & 13.9 & 8.5 & 13.4\\ 
\midrule
\multicolumn{12}{l}{\textbf{Baseline models (small batch)}} \\
\texttt{BASE} /w LS & en & 1\,k & 5\,m & 400\,k & 25.8 & 59.5 & 32.9 & 55.5 & 28.9 & 24.1 & 23.1 & 26.1 \\
\texttt{BASE} /w MLS & nl & 1.5\,k & 5\,m & 400\,k & 31.2 & 68.6 & 41.2 & 64.1 & 33.9 & 26.9 & 28.5 & 30.5 \\
\texttt{BASE} /w CV & nl & 50 & 5\,m & 400\,k & 38.8 & 72.1 & 56.3 & 76.9 & 41.4 & 33.1 & 34.4 & 34.4 \\ 
\midrule
\multicolumn{12}{l}{\textbf{Archival \texttt{nibg-pp} data}} \\
\texttt{BASE} & nl & 56\,k & 5\,m & 400\,k & 30.9 & 56.1 & 32.3 & 50.1 & 29.1 & 27.9 & 27.6 & 29.2 \\
\texttt{BASE} & nl & 56\,k & 40\,m & 400\,k & 16.7 & 34.9 & 13.8 & 25.8 & 14.3 & 17.2 & 11.4 & 15.6 \\
\texttt{LARGE} & nl & 56\,k & 40\,m & 500\,k & 13.3 & 28.5 & 9.8 & 23.2 & 11.1 & 15.0 & 8.5 & 12.9 \\
\texttt{LARGE} \cite{babu2021xlsr-large-scale} & 128 + nl & 56\,k & 80\,m & 200\,k & 11.3 & 24.1 & 7.1 & 18.1 & 9.5 & 12.7 &	6.2 & 10.6 \\
\bottomrule
\end{tabular}
\end{table*}

The fine-tuning results are shown in Table \ref{tab:segment-ft}. 
Firstly, pre-training on \texttt{w-raw} and \texttt{w-pp-1s} diverged, and are thus not shown in the table.
Secondly, \texttt{w-pp-3s} has the best overall fine-tuning performance. 
Thirdly, looking at WhisperX-based segmentation, we see that the first stage, \texttt{wx-asr}, together with the naive baseline \texttt{sequential}, have the worst fine-tuning performance (ignoring divergence).
% It is noticeable that \texttt{seq} and \texttt{wx-asr} have very similar performance, even though for \texttt{wx-asr} 13.2\,\% of the data is discarded, while \texttt{seq} only discards a marginal 0.65\,\%.
% Doing a forced alignment with a pre-trained wav2vec 2.0 model, as done with \texttt{wx-align}, leads to a large improvement in performance.
% Performing diarization (as done with \texttt{wx-diax} seems to barely affect the performance. 
% Finally, we see that the cleaning steps applied on \texttt{wx-diax} only decrease performance.  
% When pre-training with a batch size of 40 minutes (not shown in Table \ref{tab:segment-ft}, we observed a different ranking for \texttt{wx-diar} and \texttt{wx-diar-3s}. The \texttt{w-pp-3s} dataset still had the best performance, with 16.5\,\% WER when fine-tuning and evaluating on MLS. 
% However, for \texttt{wx-diar-3s}, \texttt{wx-diar}, and \texttt{seq} we respectively observed a WER of 17.1\,\%, 17.5\,\%, and 19.7\,\%. 
% When fine-tuning and evaluating on CV, we observed, respectively for \texttt{w-pp-3s}, \texttt{wx-diar-3s}, \texttt{wx-diar} and \texttt{seq}, a WER of 12.8\,\%, 14.0\,\%, 14.6\,\%, and 17.7\,\%. 
We note that performances all significantly improve when the SSL batch size was increased to 40 minuntes, indicating that all pre-processing techniques are suitable for scaling-up.
Finally, we observed that the run-time for creating the \texttt{wx-diar-3s} dataset is lower (40\,h) than \texttt{w-pp-3s} (50\,h), using a single NVIDIA A100 GPU. 

\subsection{SSL with 55\,k hours of Dutch audio data}

In the last set of experiments we scale the pre-training to the whole \texttt{nibg} dataset. 
To save time and resources, we apply the \texttt{wx-diar-3s} pre-processing technique, although based on Section \ref{exp:pre-process} we expect that the \texttt{w-pp-3s} method should have lead to slightly better performance. 
After applying the \texttt{wx-diar-3s} pre-processing on the raw 81\,k hour dataset, we obtain the pre-training dataset \texttt{nibg-pp} with 55.7\,k hours of data, having discarded 31\,\% of the raw dataset (25\,k hours). The dataset has 16.9\,M segments, with an average of 11.9 seconds, and a min and max length of respectively 3 and 30\,s.
% The length distribution peaks at 3 seconds, with \~1.8\,M utterances with a length between 3.0 and 3.1 seconds, compared to \~75\,k utterances with a length between 11.9 and 12 seconds, and \~90\,k utterances with a length between 29 and 30 seconds. 

We pre-train with the \texttt{BASE} and \texttt{LARGE} variant of wav2vec 2.0.
Weights are randomly initialized, but we also perform a run where we restart pre-training of xlsr-r-300M, a multi-lingual SSL model~\cite{babu2021xlsr-large-scale}.
We use the same configuration as detailed in Section \ref{exp:data-quality}.
However, we change the learning rate schedule to a two-stage approach as in \cite{conneau2021unsupervised} for a fairer comparison.
The two-stage schedule linearly warms-up for the first 10\,\% of total steps, the remaining 90\,\% of steps decay the LR with cosine annealing.
The initial LR is 1000 times smaller than the peak LR, while the final LR is 100 times smaller.  
For both network variants, we first perform a grid scan for the (peak) LR with a batch size of 5 minutes. 
The grid has a range of \num{1e-4} to \num{1e-3} with a step size of \num{1e-4}.
We use the best-performing LR (based on validation data) to scale to a batch of 40 minutes with the square root scaling law \cite{malladi2022scalingrule}. For a baseline, we also pre-train with English LS (960 hours), Dutch MLS (1550 hours), and Dutch CV data (50 hours).
For the baseline we use a 5 minute batch size and use the \texttt{BASE} model variant.
Fine-tuning and evaluation is done on MLS and CV as in Section \ref{exp:pre-process}, without scanning over any hyperparameters; only the initial SSL checkpoint is varied. 

The results are shown in Table \ref{tab:large-scale-ssl}.
First, we fine-tune existing pre-trained wav2vec 2.0 models \cite{baevski2020wav2vec2, conneau2021unsupervised,babu2021xlsr-large-scale} for Dutch ASR. Note that the multi-lingual models include CV~\cite{ardila2020commonvoice} and MLS~\cite{pratap2020mls} as pre-training data.
% First, we fine-tune pre-existing checkpoints, namely the \texttt{LARGE} network pre-trained with many GPUs on LS\footnote{https://huggingface.co/facebook/wav2vec2-large} in \cite{baevski2020wav2vec2}, the Dutch split of VoxPopuli\footnote{https://huggingface.co/facebook/wav2vec2-large-nl-voxpopuli} \cite{wang2021voxpopuli} in \cite{conneau2021unsupervised}, and the multi-lingual pre-training\footnote{https://huggingface.co/facebook/wav2vec2-large-xlsr-53} on the MLS, CV and BABEL, also in \cite{conneau2021unsupervised}.
We see that the English pre-training has a slightly better WER on the in-domain MLS setting.. 
For CV, the multi-lingual pre-training has the best performance. We do not see a improvement for Dutch ASR with the larger scale multi-lingual models.
When looking at our baseline experiments with a small batch size, we see that pre-training on LS, compared to Dutch MLS, or Dutch CV, is more effective. %, with significantly lower WERs on both MLS and CV evaluation data. 
We think this due the fact that the Dutch MLS dataset has very low speaker variation, and the CV dataset is very small.
When we pre-train on 55\,k hours of \texttt{nibg-pp} with the \texttt{BASE} network and the 5 minute batch size, we still see that the LS pre-training outperforms on MLS evaluation data, although for the CV data we see slightly better performance with \texttt{nibg-pp}. 
When we scale the batch size from 5 to 40 minutes, we see the WERs reduce more than 50\,\%, and performance is close to LARGE models in the first section~\cite{conneau2021unsupervised,babu2021xlsr-large-scale}. 
When we scale the network from \texttt{BASE} to \texttt{LARGE}, we see that the WERs are mostly lower compared to the multi-lingual training by \cite{conneau2021unsupervised}.
Only on the in-domain evaluation on MLS does the multi-lingual pre-training slightly outperform our mono-lingual pre-training. 
On the N-Best evaluation data, which is out-of-domain for all pre-training data, our LARGE mono-lingual model outperforms the multi-lingual models in all cases. 
We show that the performance can be increased further by continuing a multi-lingual pre-training with a (large) mono-lingual dataset, achieving 9.5\,\% WER while \cite{bualan2024systematic} reports 10\,\% for Whisper large-v2. 

\section{Conclusion}

We have shown that there are (implicit) assumptions on data quality for self-supervised learning of speech representations. 
We conclude that wav2vec 2.0 is not robust to the presence of (vocal) music.
We found that using Whisper, alongside simple heuristics, is an effective strategy to pre-process a noisy archival dataset into a qualitative pre-training dataset.
We confirm that mono-lingual pre-training increases performance compared to multi-lingual pre-training, more so on out-of-domain data.

\section{Acknowledgment}

This work was sponsored by NWO - Domain Science for the use of supercomputer facilities  (Snellius, \url{www.surf.nl}).

\bibliographystyle{IEEEtran}
\bibliography{mybib}

\end{document}